\documentclass[twocolumn]{aastex63}
\usepackage{mathtools}
\usepackage{hyperref}
\submitjournal{The Astronomical Journal \\
\href{https://doi.org/10.3847/1538-3881/abf045}{DOI:10.3847/1538-3881/abf045}}

\shorttitle{\href{https://doi.org/10.3847/1538-3881/abf045}{DOI:10.3847/1538-3881/abf045}}
\shortauthors{Hao et al.}

\begin{document}

\title{Effect of near-field distribution on transmission characteristics of fiber-fed Fabry--Perot etalons}

\correspondingauthor{Huiqi Ye}
\email{hqye@niaot.ac.cn}

\author[0000-0002-9259-6176]{Jun Hao}
\affiliation{National Astronomical Observatories/Nanjing Institute of Astronomical Optics \& Technology, Chinese Academy of Sciences \\
Nanjing 210042, People's Republic of China}
\affiliation{CAS Key Laboratory of Astronomical Optics Technology, Nanjing Institute of Astronomical Optics \& Technology \\
Nanjing 210042, People's Republic of China}
\affiliation{University of Chinese Academy of Sciences \\
Beijing 100049, People's Republic of China}

\author{Liang Tang}
\affiliation{National Astronomical Observatories/Nanjing Institute of Astronomical Optics \& Technology, Chinese Academy of Sciences \\
Nanjing 210042, People's Republic of China}
\affiliation{CAS Key Laboratory of Astronomical Optics Technology, Nanjing Institute of Astronomical Optics \& Technology \\
Nanjing 210042, People's Republic of China}

\author[0000-0002-7216-0412]{Huiqi Ye}
\affiliation{National Astronomical Observatories/Nanjing Institute of Astronomical Optics \& Technology, Chinese Academy of Sciences \\
Nanjing 210042, People's Republic of China}
\affiliation{CAS Key Laboratory of Astronomical Optics Technology, Nanjing Institute of Astronomical Optics \& Technology \\
Nanjing 210042, People's Republic of China}

\author{Zhibo Hao}
\affiliation{National Astronomical Observatories/Nanjing Institute of Astronomical Optics \& Technology, Chinese Academy of Sciences \\
Nanjing 210042, People's Republic of China}
\affiliation{CAS Key Laboratory of Astronomical Optics Technology, Nanjing Institute of Astronomical Optics \& Technology \\
Nanjing 210042, People's Republic of China}
\affiliation{University of Chinese Academy of Sciences \\
Beijing 100049, People's Republic of China}

\author{Jian Han}
\affiliation{National Astronomical Observatories/Nanjing Institute of Astronomical Optics \& Technology, Chinese Academy of Sciences \\
Nanjing 210042, People's Republic of China}
\affiliation{CAS Key Laboratory of Astronomical Optics Technology, Nanjing Institute of Astronomical Optics \& Technology \\
Nanjing 210042, People's Republic of China}

\author{Yang Zhai}
\affiliation{National Astronomical Observatories/Nanjing Institute of Astronomical Optics \& Technology, Chinese Academy of Sciences \\
Nanjing 210042, People's Republic of China}
\affiliation{CAS Key Laboratory of Astronomical Optics Technology, Nanjing Institute of Astronomical Optics \& Technology \\
Nanjing 210042, People's Republic of China}

\author{Kai Zhang}
\affiliation{National Astronomical Observatories/Nanjing Institute of Astronomical Optics \& Technology, Chinese Academy of Sciences \\
Nanjing 210042, People's Republic of China}
\affiliation{CAS Key Laboratory of Astronomical Optics Technology, Nanjing Institute of Astronomical Optics \& Technology \\
Nanjing 210042, People's Republic of China}

\author{Ruyi Wei}
\affiliation{Xi'an Institute of Optics \& Precision Mechanics, Chinese Academy of Sciences \\
 Xi'an 710119, People's Republic of China}

\author{Dong Xiao}
\affiliation{National Astronomical Observatories/Nanjing Institute of Astronomical Optics \& Technology, Chinese Academy of Sciences \\
	Nanjing 210042, People's Republic of China}
\affiliation{CAS Key Laboratory of Astronomical Optics Technology, Nanjing Institute of Astronomical Optics \& Technology \\
	Nanjing 210042, People's Republic of China}



\begin{abstract}

Fiber-fed etalons are widely employed in advanced interferometric instruments such as gravitational-wave detectors, ultrastable lasers and calibration reference for high-precision spectrographs.  We demonstrate that variation in near-field distribution of the feeding fiber would deteriorate the spectrum precision of the fiber-fed Fabry--Perot etalon, especially when precision at the order of $\rm 3 \times 10^{-10}$ or higher is required.  The octagonal fiber reinforced with double scrambler could greatly improve the steadiness and uniformness of the near-field distribution. When building wavelength calibrators for sub-m $\rm s^{-1}$ precision radial-velocity instruments, the double scrambler should be considered meticulously.

\end{abstract}



\section{Introduction} \label{sec:intro}

Fabry--Perot etalons (FPEs) are used in a variety of advanced applications, such as gravitational-wave detection \citep{miller2012control}, metrology \citep{chen2019absolute} and laser stabilization \citep{alnis2008subhertz}. For astronomical observations, besides its long history in solar imaging and scanning spectroscopy, FPEs have emerged as a new type of economical and potentially reliable spectral reference for simultaneous calibration of high-precision astronomical spectrographs. Since the discovery of the first exoplanet orbiting around a solar-type star \citep{mayor1995jupiter}, the radial-velocity (RV) method has become one of the most important techniques for exoplanet search and characterization. It measures the subtle change in Doppler shift induced by the exoplanet candidate on the host star spectrum along our line of sight. To detect an Earth-like exoplanet, a long-term precision of $\rm 10~cm~s^{-1}$ (corresponding to a fractional stability better than $\rm 3 \times 10^{-10}$) or higher would be required. This ever-increasing demand for high precision is pushing for new calibration techniques in the stead of traditionally used hollow-cathode lamps (HCLs) \citep{lovis2006exoplanet,lovis2007new} or gas absorption cells \citep{wang2019calibrating} in order to achieve sub-m $\rm s^{-1}$ long-term stability.

Laser frequency combs (LFCs; \citealt{steinmetz2008laser, wilken2010high}) have been demonstrated to be excellent calibration sources for next-generation ultraprecision RV instruments. However, the high price, complex structure, and high-maintenance demand has made them hard to access for small-scale projects. In view of this, passively stabilized FPE illuminated with a broadband source has been developed as an economical alternative \citep{wildi2010fabry}. For which, stabilities of $\rm 10~cm~s^{-1}$ during one night and $\rm 1~m~s^{-1}$ over 60 days have been demonstrated on the HARPS and HARPS-N spectrographs \citep{wildi2012passive}. Ideally, the transmitted peaks are equally spaced in the frequency domain and cover the entire spectral span of the calibrated instrument. The line width and spacing of the peaks are defined by the finesse and free spectral range (FSR) of the etalon. By anchoring the FPE peaks to atomic transition \citep{reiners2014laser,schwab2015stabilizing}, it has been shown that a high tracking/locking-precision at the $\rm cm~s^{-1}$ level can be reached by this type of calibrator.

To achieve highly stable passive environmental control, high efficiency, and better compatibility with fiber-fed spectrographs, FPE-based calibrators are usually fed by a multimode fiber (MMF). \citet{cersullo2017new} discussed the influence of fiber diameter and decentering on the spectral characteristics of FPE-based calibrators and showed that the line position and shape stability of the transmitted peaks are significantly influenced by the alignment condition of the input fiber. Yet, the influence of nonuniform near-field intensity distribution of the fiber output has not been fully addressed. Here, we analyze the effects of near-field distribution of the input fiber on the transmission characteristic of FPEs through numerical simulation. We then propose and experimentally demonstrate a mitigation method that combines the use of octagonal fiber and a double scrambler, an optical relay that exchanges the near and far fields \citep{hunter1992scrambling, halverson2015efficient, ye2020experimental}, by which both the near-field and far-field illumination patterns are stabilized.

\section{Theory of the fiber-fed Fabry--Perot etalon} \label{sec:theo}

The FPE consists of two flat parallel mirrors coated with a highly reflective coating on the inner surfaces. Suppose an FPE is illuminated with a perfectly collimated beam. The frequency at peak transmission is an integer multiple of the FSR. In the frequency domain, the FSR is given by

\begin{equation}
\Delta \nu = \frac{c}{2 n d \cos \theta},
\end{equation}

where $d$ denotes the distance between the two mirrors, $n$ is the refractive index of the gap medium, and $\theta$ is the incident angle of the input beam. The finesse $\mathcal{F}$ of the etalon is defined as the ratio of FSR to the FWHM of the transmitted peak.

\begin{figure}[ht!]
	\centering
	\includegraphics[width=8cm]{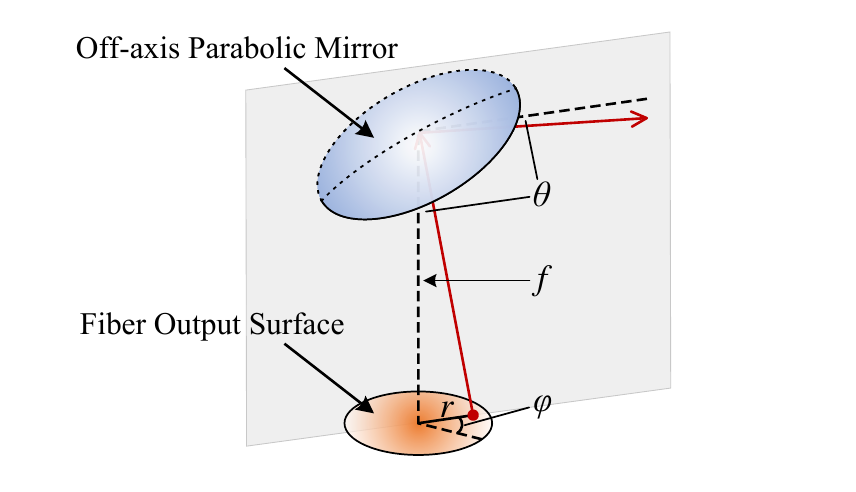}
	\caption{Diagram of the collimated light beam. Light from different locations on the fiber surface results in different incidence angles. $f$ is the reflected focal length(RFL).
		\label{fig1}}
\end{figure}

Assuming that the absorption and scattering losses are negligible, the transmission function of the FPE is given by

\begin{equation}
\tau(\lambda, \theta) = \frac{1}{1 + F \sin ^2 (\delta(\lambda, \theta) / 2)},
\end{equation}

in which $\delta$ is the round-trip phase shift

\begin{equation}
	\delta(\lambda, \theta) = \frac{4\pi}{\lambda}n d \cos \theta,
\end{equation}

and $F$ is the finesse coefficient, defined as

\begin{equation}
F = \frac{4 \rho}{(1 - \rho)^2},
\end{equation}

with $\rho$ as the reflectivity of the coatings. Assume achromatic reflective collimator with off-axis parabolic mirror is used to collimate the light beams. Figure \ref{fig1}  shows the diagram of the collimated light beam. Due to the fiber having an extended surface at the output instead of an ideal point, the off-axis light will incident onto the etalon with a small angle after collimating.

Suppose the intensity at a given point on the fiber cross-section is expressed as $p (r, \varphi)$ in polar coordinates, with the fiber center as the origin. According to the geometrical relation and properties of an off-axis parabolic mirror, the incident angle of the ray coming from this point after collimation is

\begin{equation}
\theta(r) = \arctan  \frac{r}{f},
\end{equation}

where $f$ is the reflected focal length (RFL) of the reflective collimator. The effective transmission function at point $p (r, \varphi)$ can be expressed as

\begin{equation}
\tau(\lambda, r) = \frac{1}{1 + F \sin ^2 (2 n d \cos (\arctan (r / f))/ \lambda)}.
\end{equation}

Thus, the effective transmission function $I_{tran}(\lambda)$ of a fiber-fed FPE is an integral over the entire fiber output surface:

\begin{equation}
I_{tran}(\lambda) = \iint p (r, \varphi) \tau(\lambda, r) r dr d\varphi.
\end{equation}

\begin{figure}[ht!]
	\centering
	\includegraphics[width=9cm]{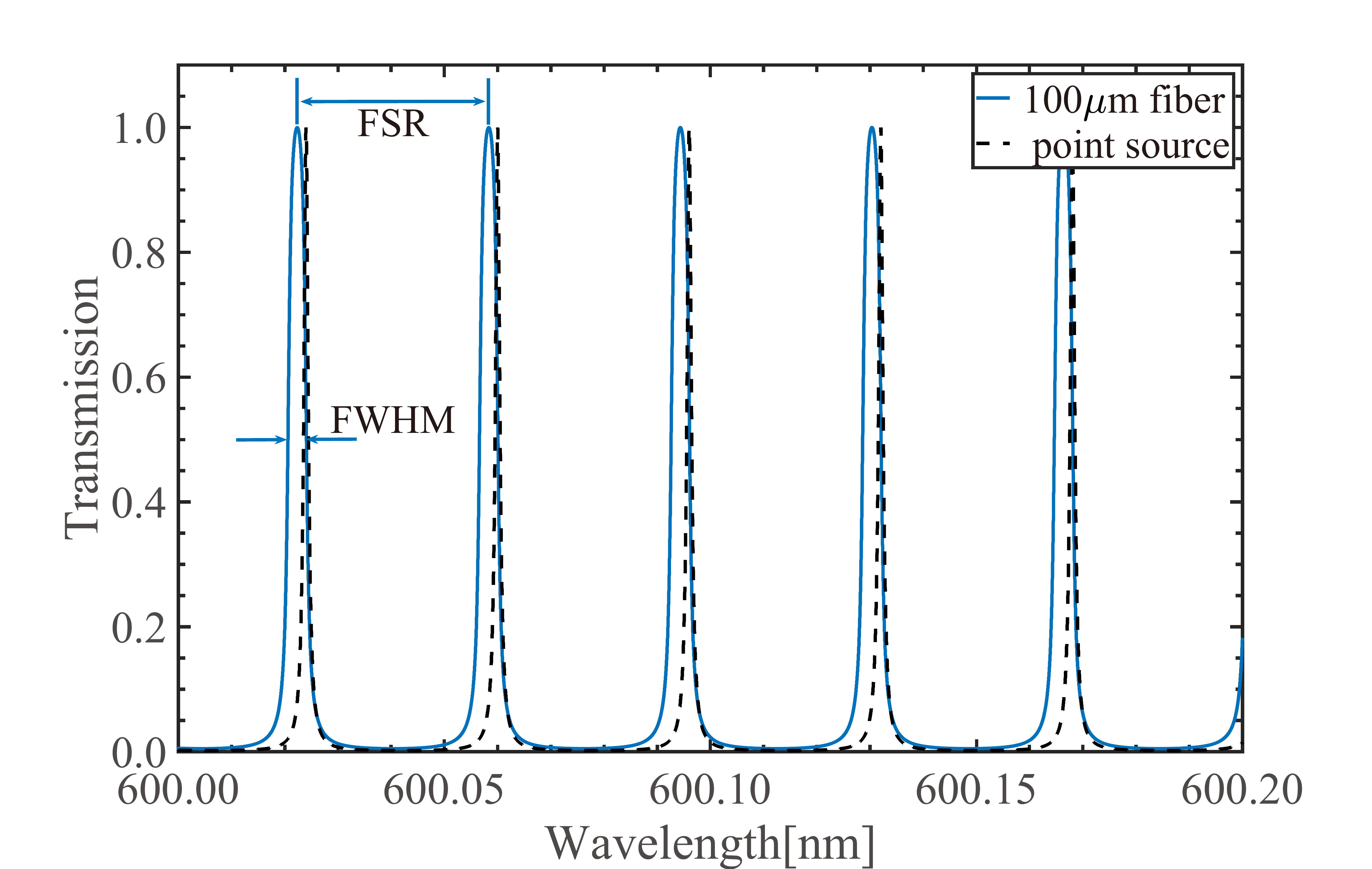}
	\caption{Simulation of FPE transmission spectra.
		\label{fig2}}
\end{figure}

\begin{figure}[ht!]
	\centering
	\includegraphics[width=8.5cm]{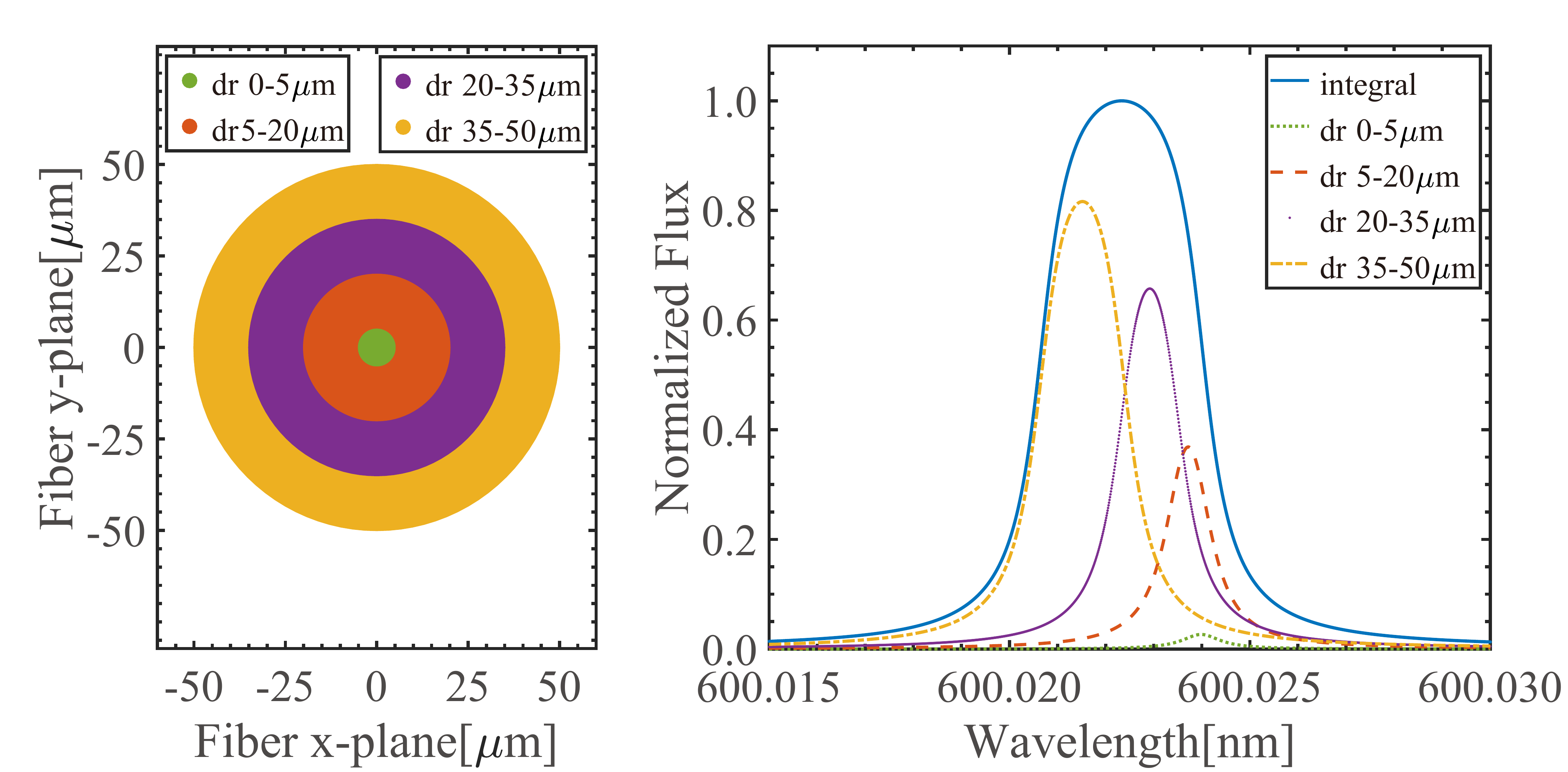}
	\caption{Anatomic diagram of the contribution of light from different near-field radius to the transmission spectrum. Left panel shows the individual rings $dr$ for different areas, and the corresponding transmission spectrum is shown in the right panel. The integral spectrum is the superposition of curves formed by individual rings $dr$.
		\label{fig3}}
\end{figure}

For simulation, an FPE and reflective collimator with properties as listed in Table \ref{table1} are employed. Figure \ref{fig2} shows the theoretical transmission function of the FPE with a fiber of $\rm 100 ~\mu m$ diameter at normal incidence (solid line). For comparison, the case of ideal on-axis transmission (dashed line) is also shown. The spectral line generated by the extended source is broadened compared with the point source. An FWHM $w_0$ of $\rm 1.24 \times 10^{-3} nm$ is calculated for the ideal case. While FWHM $w_f$ of $\rm 3.29 \times 10^{-3} nm$ and an effective finesse of 10.94 are results in the $\rm 100 ~\mu m$ case, much lower than the specified value. The reason for such broadening can be explained as follows: the spectrum formed by the entire fiber surface is the superposition of segments formed by individual elements $dr$, as shown in Figure \ref{fig3}. An element further away from the fiber center results in a larger incident angle, thus contributing to the increase of blue-shifted components of the integrated curve. 

\begin{table}[!htb]
	\centering
	\caption{Simulation Parameters \label{table1}}
	{\begin{tabular}{lcl}
			\hline
			\hline
			Parameter & Symbol & Value \\ \hline
			FPE gap & $d$ & 5 mm \\
			FSR & $\Delta \nu$ & 30 GHz (0.036 nm @ 600 nm) \\ 
			Finesse & $\mathcal{F}$ & 37 \\
			Reflectivity & $\rho$ &  0.92 \\
			RFL & $f$ & 15 mm \\
			Refractive index & $n$ & 1 \\
			Fiber core diameter & $d_f$ & $\rm 100 ~\mu m$ \\
			\hline
	\end{tabular}}
\end{table}

\section{Effect of fiber near-field distribution on the FPE transmission} \label{sec:effec}

For radial-velocity measurements, the utilization of fiber input decouples the spectrograph from the telescope focus, significantly improving the instrumental profile stability. However, the near-field distribution of an optical fiber is sensitive to the injection conditions and external disturbances. Variance of illumination, movement, and temperature fluctuation of the feeding fiber will result in changes at the fiber output and image shifts on the detector, manifesting as false RV shifts \citep{avila2008optical, brown1991detection, halverson2015efficient}. For coherent calibration light sources, such as LFCs, such a phenomenon is more prominent due to mode interference and speckle issues \citep{mahadevan2014suppression}. For FPE-based calibrators, an incoherent broadband light source can be employed; however, the influence of near-field distribution variation on the calibration precision is yet to be determined.

\begin{figure}[ht!]
	\centering
	\includegraphics[width=8.5cm]{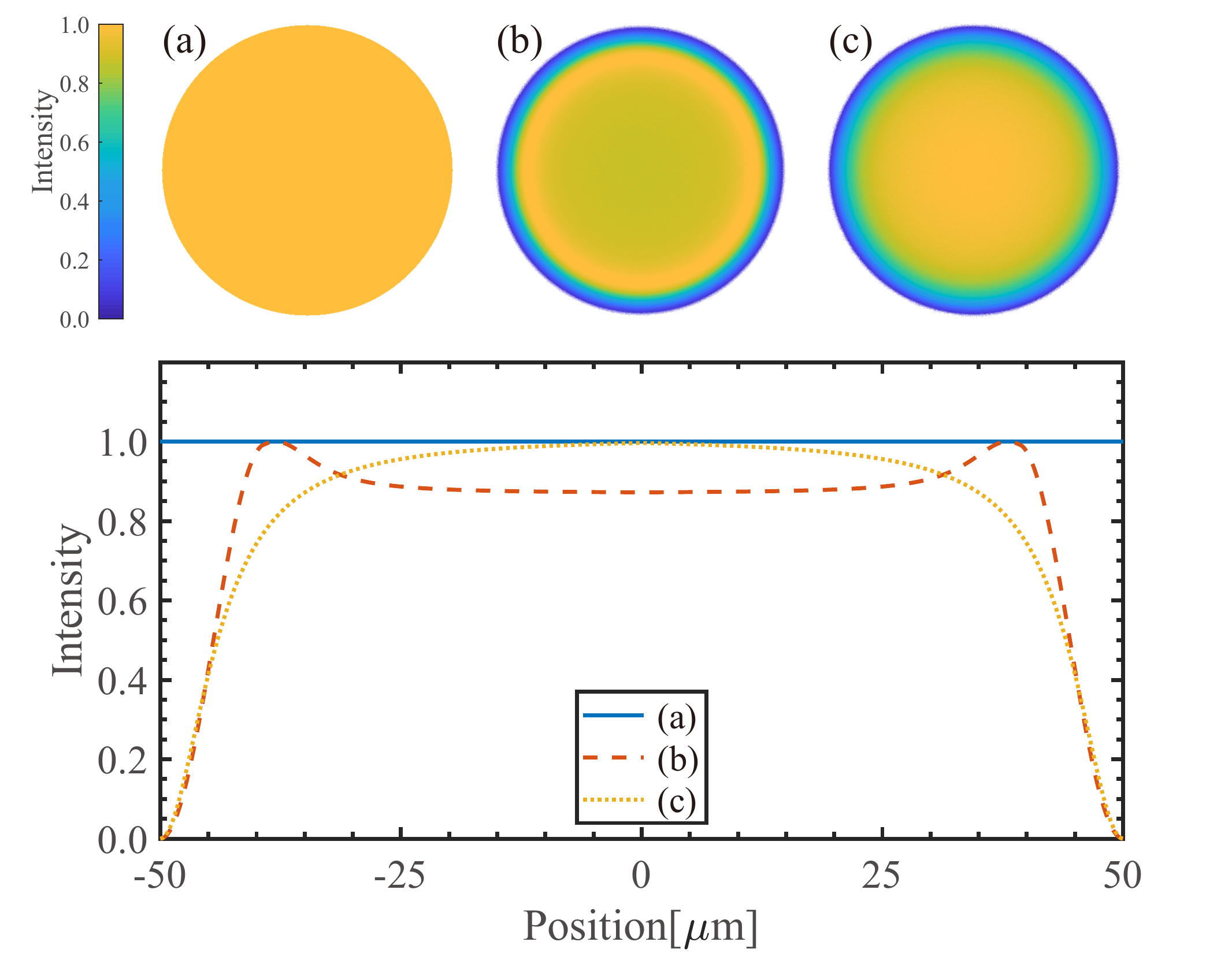}
	\caption{Three simulated intensity distributions of the near-field (core diameter $\rm = 100  ~\mu m$). Top panels show the diagram of the fiber cross-section. Bottom panel shows the 1D relative intensity distribution across the center of the fiber. Panel (a) is the ideal case, panel (b) may occur when the light source is injected off-center, and panel (c) may occur when it is injected on-center.
		\label{fig4}}
\end{figure}

Assume that the fiber is well coaxially aligned with the collimator. It has been previously demonstrated in the literature \citep{heacox1986application, heacox1987radial, chen2016simulation} that the radial distribution of near-field intensity distribution of a circular MMF can be uneven while the azimuthal distribution is generally uniform under different injection conditions, as shown simulatively in Figure \ref{fig4}. Figure \ref{fig4}(a) is the ideal case where intensity variation is ignored. This ideal model served as the framework of the above FPE line-broadening analysis, but is unattainable in reality. The actual intensity distribution will be similar to Figure \ref{fig4}(b) (off-center injection) or Figure \ref{fig4}(c) (on-center injection).

\begin{figure}[ht!]
	\centering
	\includegraphics[width=8.5cm]{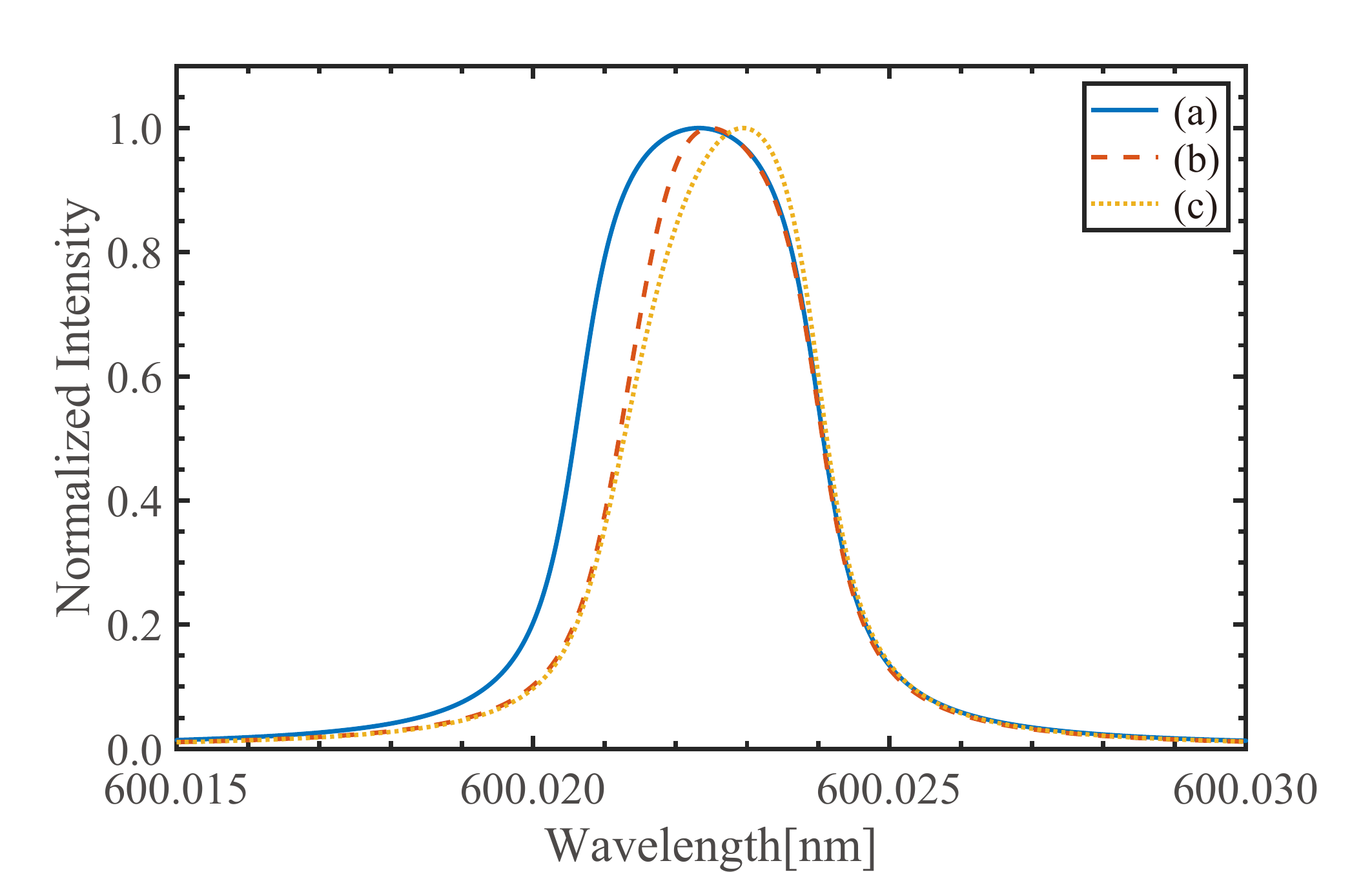}
	\caption{The intensity-normalized spectra resulting from different near-field distributions in Figure \ref{fig4}.  \label{fig5}}
\end{figure}

These different intensity distributions are imported separately into the FPE simulation model, and the normalized transmission spectra are shown in Figure \ref{fig5}. As the radial intensity distribution is not uniform, light on the edge of the fiber surface contributes less to the final curve compared to the ideal case, shifting the peak to the red end. Due to this phenomenon, the symmetry of the peak is broken.  The Gaussian fitted results of spectral drifts and equivalent RV shifts are shown in Table \ref{table2}.

\begin{table}[!htb]
	\centering
	\caption{Fitted Values of the Simulated Spectra with Gaussian Functions \label{table2}}
	{\begin{tabular}{llll}
			\hline
			\hline
			 & FWHM & $\Delta \lambda / \lambda_{(a)}$ & Equivalent RV \\ \hline
			Figure 5(a) & $\rm 3.29 \times 10^{-3} nm$ & 0 & 0 \\ 
			Figure 5(b) & $\rm 2.28 \times 10^{-3} nm$ & $\rm 4.98 \times 10^{-7}$ & $\rm 149.44~m~s^{-1}$ \\
			Figure 5(c) & $\rm 2.81 \times 10^{-3} nm$ & $\rm 6.86 \times 10^{-7}$ & $\rm 205.90~m~s^{-1}$ \\
			\hline
	\end{tabular}}
\end{table}

By comparing the fitting results of Figure \ref{fig5}(a)--(c), it can be seen that a very large spectral drift will occur when the near-field intensity distribution of the fiber goes through dramatic change.

\begin{figure}[ht!]
	\centering
	\includegraphics[width=8.5cm]{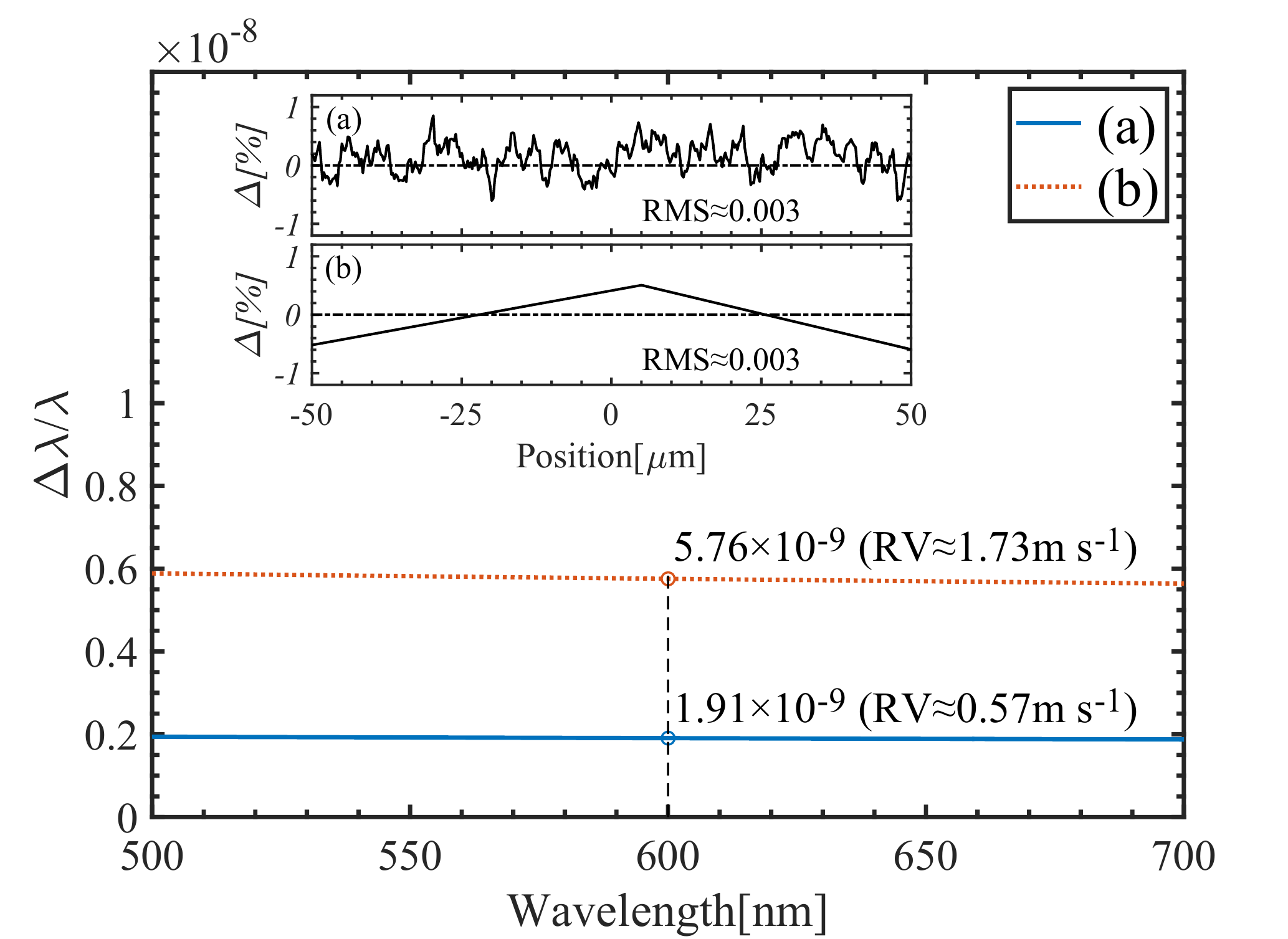}
	\caption{Wavelength dependency of the fractional spectral shift induced by two slight intensity fluctuations as shown in the inset. Inset: (a) could happen when the system is slightly jitter (such as fiber movement) and (b) may occur when the point of incidence changes (from the edge to the center of the light source).  \label{fig6}}
\end{figure}

The question remains: How stable does the near-field intensity distribution of the fiber need to be in order to achieve calibration stability at the $\rm 10~cm~s^{-1}$ level? To quantify the effect of turbulence, two different types of slight intensity fluctuations of less than $\pm 0.5 \%$ (as shown in Figure \ref{fig6} inset) are introduced into the case of Figure \ref{fig4}(a). The spectral lines obtained by superimposing Figure \ref{fig4}(a) with these two different intensity fluctuations are then compared with the result from Figure \ref{fig4}(a), respectively. Consider a calibrator of a spectrograph that operates within the wavelength range of 500--700 nm, the spectral drifts of the FPE peaks within that operation range are calculated and plotted in Figure 6. The two different fluctuations result in different wavelength dependency of the induced drifts, with maximum variations of $3.5\%$ and $4.3\%$, respectively, between the blue and red end, both minute in nature. For direct comparison, the amplitudes of induced spectral drifts around 600 nm are $\rm 1.91 \times 10^{-9}$ and $\rm 5.76 \times 10^{-9}$ (equal to $\rm 0.57~m~s^{-1}$ and $\rm 1.73~m~s^{-1}$ of RV shift), for the two different fluctuations. Errors at this level are unneglectable and could potentially affect the calibration precision, hence the need for special measures to ensure the system stability.

\section{Experimental demonstration of the effect of near-field distribution disturbance} \label{sec:experi}

\begin{figure}[ht!]
	\centering
	\includegraphics[width=8.5cm]{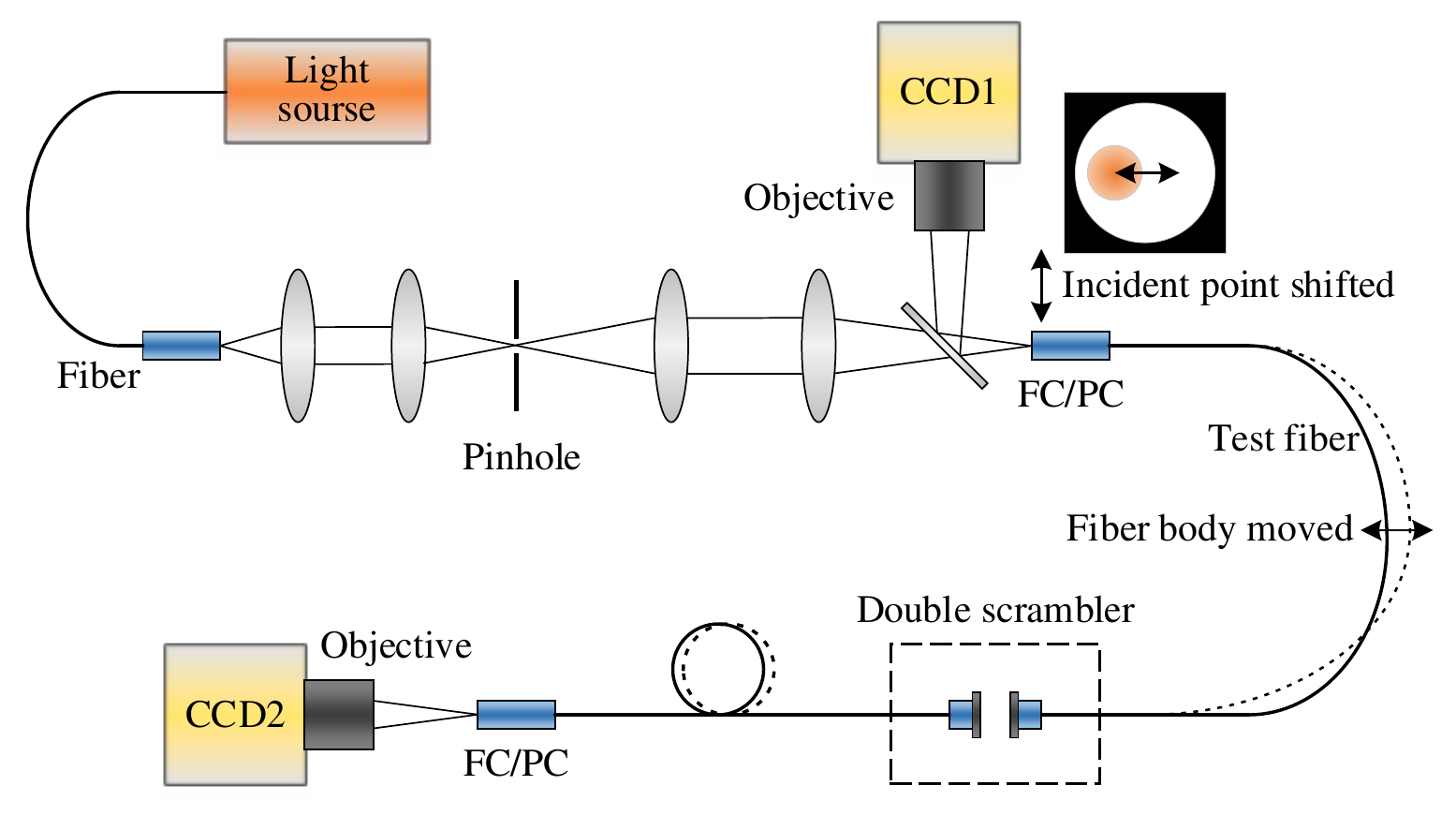}
	\caption{Schematic diagram of the experimental setup used for near-field characterization. 
		 \label{fig7}}
\end{figure}

\begin{figure}[ht!]
	\centering
	\includegraphics[width=8cm]{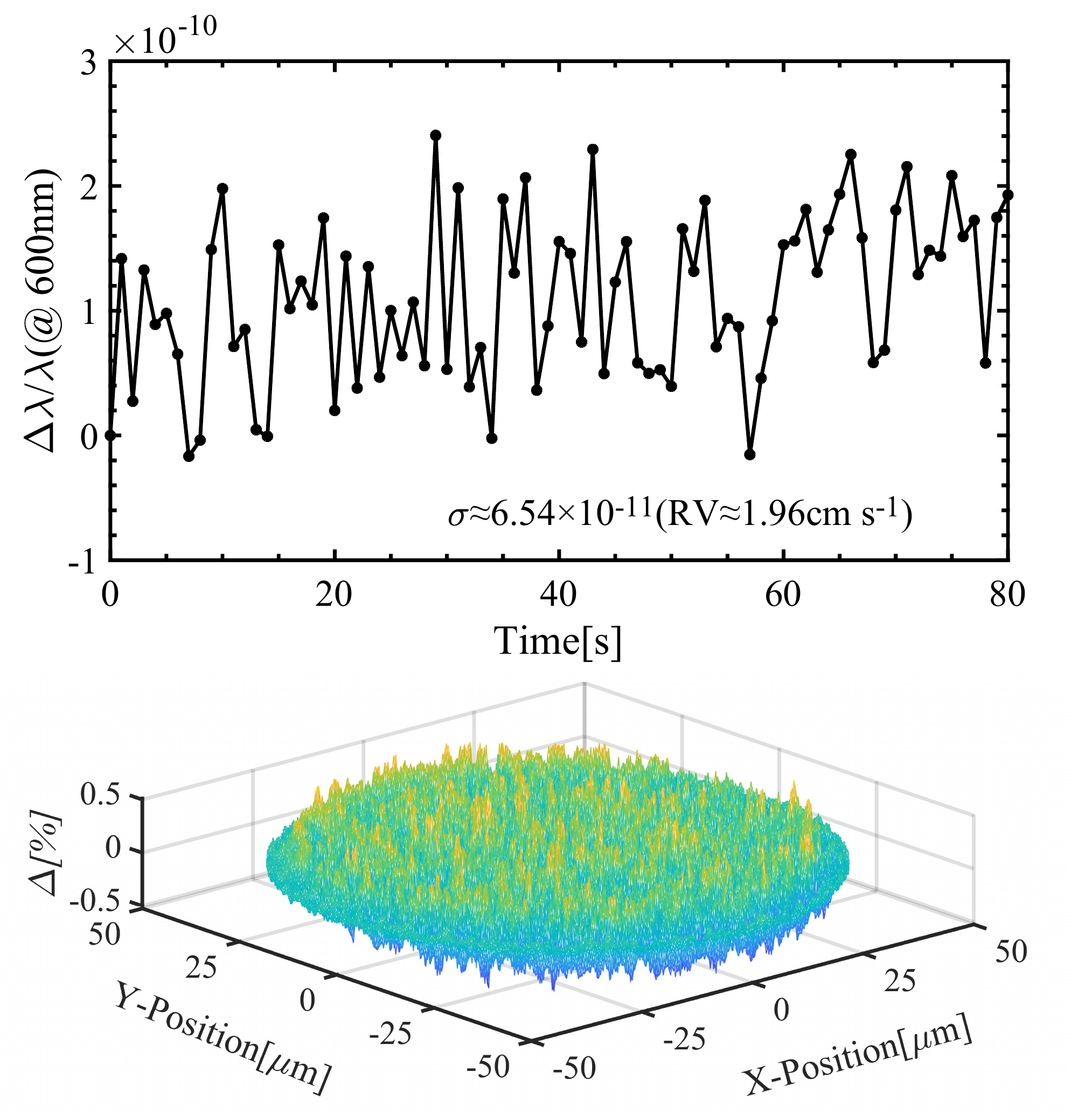}
	\caption{Stability test of experimental setup. Continuous measurements of the circular fiber near-field in the undisturbed condition with 1s interval time over 80s. Upper panel shows the fractional displacement ($\Delta\lambda / \lambda$) of the transmission peaks corresponding to initial time calculated by importing each pixel into the FPE simulation model, and the standard deviation ($\sigma$) is marked in the graph. Bottom panel shows the difference ($\Delta$, in percentage) between the two sets of near-field intensity distribution that produced the highest and lowest drift values.
		\label{fig8}}
\end{figure}

In order to study the impact of the intensity jitter of the optical fiber, the experimental setup shown in Figure \ref{fig7} is used to measure the near-field intensity distribution on the fiber output surface. The light of a broadband source (halogen lamp, 500-700 nm) is led into the setup by a $\rm 400 ~\mu m$ optical fiber, the endface of which is imaged at the $\rm 50 ~\mu m$ pinhole, then projects onto the test fiber. The near-field intensity distribution of the test fiber is then obtained by CCD with a microscope. Before experiment, the stability of setup should be demonstrated. The intensity distribution of the optical fiber output end is measured without any disturbance with 1s interval time over 80s. As we can see from Figure \ref{fig8}, the fractional displacement ($\Delta\lambda / \lambda$) of the transmission peaks obtained by substituting the sets of measurements into the FPE simulation model pixel by pixel is approximately $\rm 6.54 \times 10^{-11}$ (equal to $\rm 1.96~cm~s^{-1}$ of RV shift).

\begin{figure}[ht!]
	\centering
	\includegraphics[width=8.5cm]{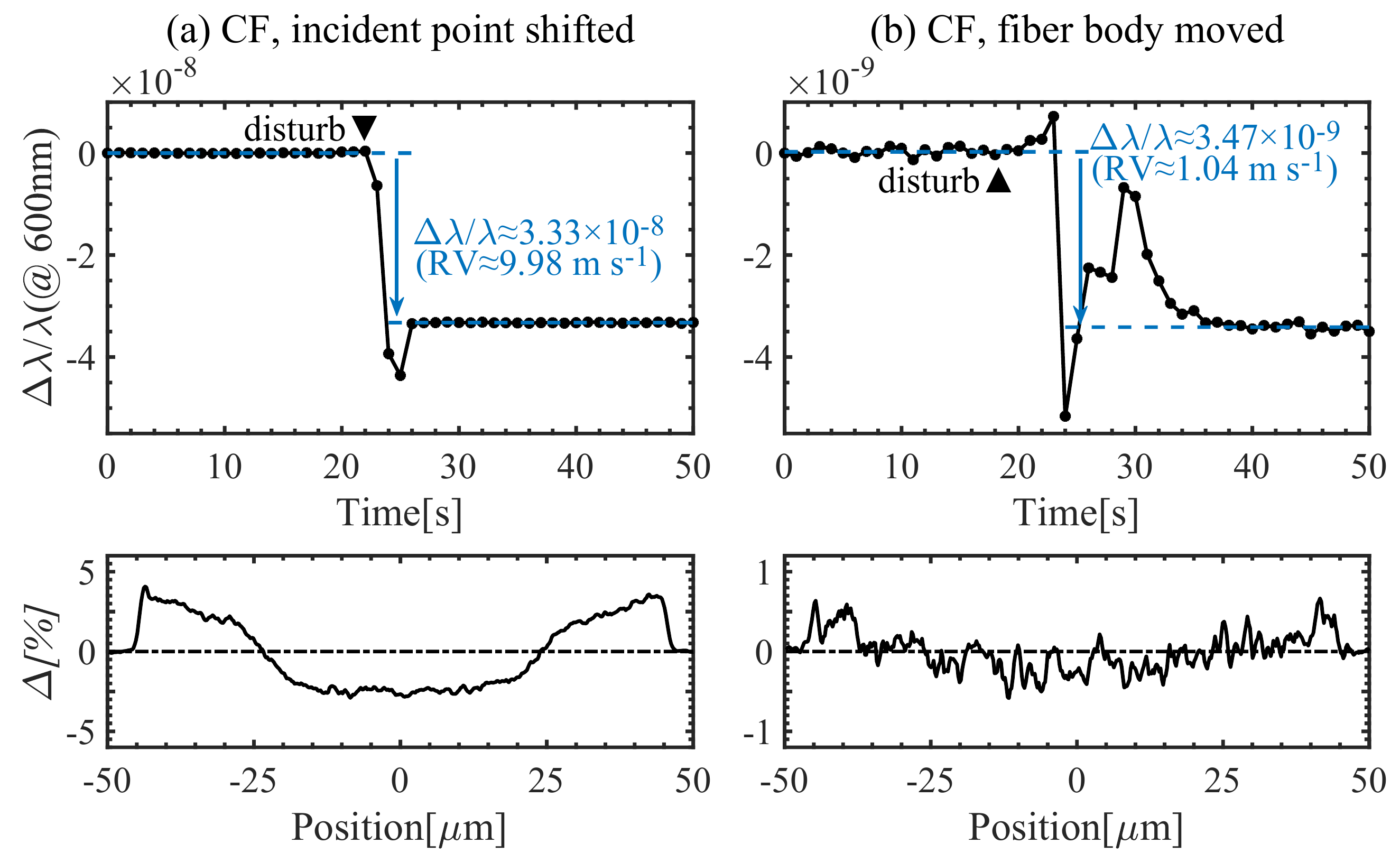}
	\caption{Two sets of near-field measurements of different circular fiber (CF) configurations. (a) The incident point is shifted off-center by $\rm 20 ~\mu m$; (b) the fiber body is moved to a different position. Upper panels show the fractional displacement ($\Delta\lambda / \lambda$), the dashed lines show the average of the first or last 10 data, and the drifts caused by the disturbances are marked in the graphs. Bottom panels show the difference ($\Delta$, in percentage) of 1D normalized intensity distribution across the center of the fiber before and after disturbance, calculated by each taking the average of the first 10 near-field frames and the last 10 frames.
		\label{fig9}}
\end{figure}

\begin{figure*}[ht!]
	\centering
	\includegraphics[width=18cm]{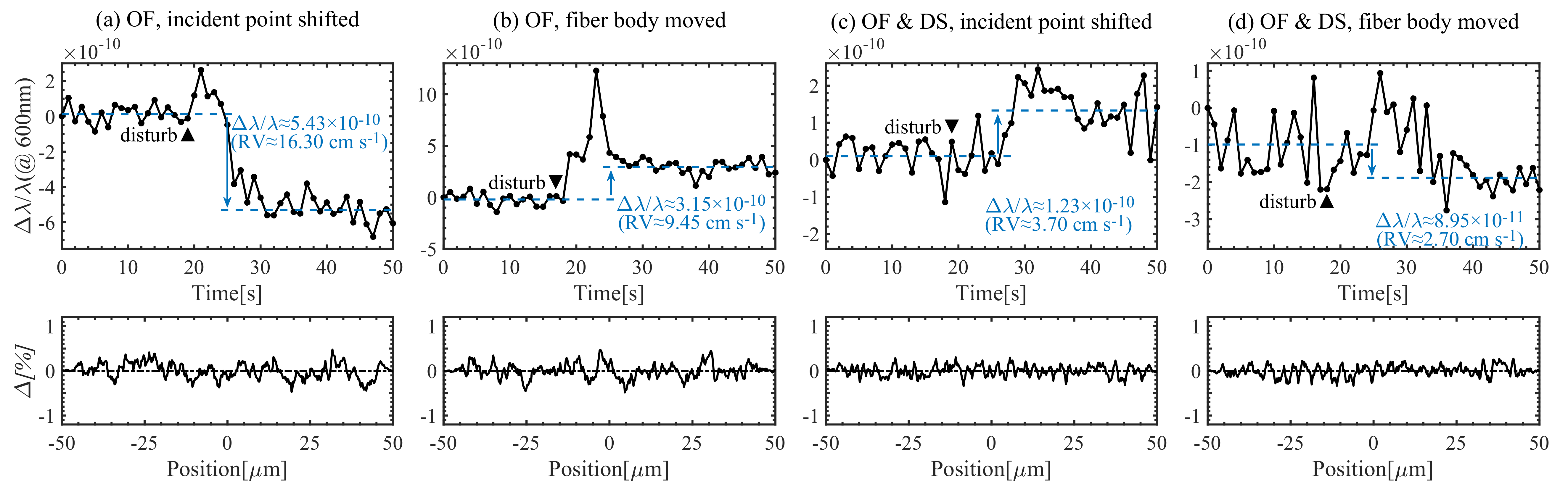}
	\caption{Four sets of near-field measurements of different octagonal fiber (OF) configurations. (a) the incident point is shifted off-center; (b) the fiber body is moved to a different position; (c) the incident point is shifted off-center while a double scrambler (DS) is used; (d) the fiber body is moved to a different position, while a double scrambler is used.  \label{fig10}}
\end{figure*}

With the fiber near-field being kept on catch, the change of a single test parameter (incident point shift or fiber body movement) is introduced. It is carefully ensured that the time elapsed of the whole process does not exceed 80s. In such a way, the measurements contain information regarding both the stability of setup and the changed variable. In Figure \ref{fig9}, the results of circular fiber (CF, 5 m in length) with different disturbances are shown. The incident point and state of fiber are altered to explore the influence of possible disturbances, as shown in Figure \ref{fig7}. In the first test, the incident point shifts off-center by $\rm 20 ~\mu m$, which causes a fractional drift of $\Delta\lambda / \lambda \approx 3.33 \times 10^{-8}$ ($\rm RV \approx 9.98~m~s^{-1}$). Another test is done by moving the body of the fiber to a different position, and the fractional drift caused is approximately by $\Delta\lambda / \lambda \approx 3.47 \times 10^{-9}$ ($\rm RV \approx 1.04~m~s^{-1}$). This order of magnitude is much higher than in the undisturbed condition. It is evident that the intensity distribution in the near-field of the CF is sensitive to environment disturbances. Therefore, to maintain high calibration precision, the fiber needs to be placed in an absolutely stable condition, which is not always feasible in practical environments.

A number of studies have shown that double scrambler can improve the stability of near-field intensity distribution \citep{hunter1992scrambling, halverson2015efficient}. Additionally, it has been shown that octagonal fibers demonstrate more uniform and stable near-field output compared to CFs \citep{chazelas2010new}. Here, a combination of double scrambler and octagonal fiber \citep{ye2020experimental} as a solution to meet the above discussed precision requirement is proposed and discussed.

Switching to an octagonal fiber (3 m in length), we perform similar tests as done for the CF. The results are shown in Figure \ref{fig10}. In which the way of intensity redistribution in the near-field after disturbance may vary across different fiber configurations, resulting in a different sign of the fractional displacement ($\Delta\lambda / \lambda$), but the effect is equivalent. When the point of incidence shifts, it causes a fractional drift of $\Delta\lambda / \lambda \approx 5.43 \times 10^{-10}$ ($\rm RV \approx 16.30~cm~s^{-1}$) in the case of an octagonal fiber, a significant improvement compared to the similar case in CF. By adding a double scrambler operating during the shift (the fiber length before and after the double scrambler is 3 m and 3 m, respectively), the fractional drift is further reduced to $1.23 \times 10^{-10}$ ($\rm RV \approx 3.70~cm~s^{-1}$). Similarly, when the fiber body is moved to a different position, a drift of  $\Delta\lambda / \lambda \approx 3.15 \times 10^{-10}$ ($\rm RV \approx 9.45~cm~s^{-1}$) is observed. The employment of the double scrambler drops the value to $8.95 \times 10^{-11}$ ($\rm RV \approx 2.70~cm~s^{-1}$), almost the limit of setup. The octagonal fiber performs satisfactorily, and the coemployment of a double scrambler excellently improves the performance.

The experimental results show that the octagonal fiber along with a double scrambler can significantly reduce the fractional drift caused by near-field intensity distribution in a fiber-fed FPE, making it less sensitive to environmental disturbances and more suitable for use as a high-precision calibrator.

\section{Summary} \label{sec:summa}

The intensity  distribution variation in the near-field of the CF output will result in displacement and distortion of the transmitted spectral peaks of the FPE. This in turn leads to errors in applications where extreme high precision is required. Even a mere $\pm 0.5 \%$ of intensity fluctuation can cause a corresponding ~$\rm \sim 3 \times 10^{-9}$ ($\rm 1~m~s^{-1}$) fractional shift of the transmission peaks in an FPE-based calibrator, a considerable error in the study of exoplanet through the RV method, thus putting rigorous requirement on the operation environment. This work proposes and proves via numerical simulations that octagonal fiber with double scrambler can substantially increase the stability of output near-field intensity distribution. The spectral peak deviation caused by intensity fluctuation can be reduced to $\rm  < 1.5 \times 10^{-10}$ ($\rm 5~cm~s^{-1}$), demonstrating outstanding resistance to environmental disturbances.

\acknowledgments

This work was partially supported by the National Natural Science Foundation of China (grant Nos. 11773044, 11727806, 11673046, 11873071, and 11903060) and the Operation, Maintenance and Upgrading Fund for Astronomical Telescopes and Facility Instruments, budgeted from the Ministry of Finance of China (MOF) and administrated by the Chinese Academy of Sciences (CAS). L.T. acknowledges support from the China Postdoctoral Science Foundation (2020M671638) and Jiangsu Planned Projects for Postdoctoral Research Funds.

\bibliography{REFER}{}
\bibliographystyle{aasjournal}



\end{document}